\documentclass[Letter, longauth]{aa}
\usepackage{graphicx}
\usepackage{txfonts}
\usepackage{natbib}
\usepackage[colorlinks=true,citecolor=blue,breaklinks=true]{hyperref}
\usepackage{xspace}
\usepackage{amsmath,amssymb}
\usepackage{gensymb}
\usepackage{enumitem}
\usepackage{upgreek}
\usepackage{subcaption}
\usepackage{tabularx}
\usepackage{todonotes}
\usepackage{afterpage}
\usepackage{placeins}
\usepackage{xcolor}
\usepackage[page]{appendix}
\usepackage{caption}
\usepackage{float}

\bibpunct{(}{)}{;}{a}{}{,}

\newcommand{\MJup}{\ensuremath{M_{\mathrm{Jup}}}\xspace}
\newcommand{\RJup}{\ensuremath{R_{\mathrm{Jup}}}\xspace}
\newcommand{\Teff}{\ensuremath{T_{\mathrm{e\!f\!f}}}\xspace}
\newcommand{\logg}{\ensuremath{\log g}\xspace}
\newcommand{\met}{\ensuremath{\mathrm{[Fe/H]}}\xspace}
\newcommand{\co}{\ensuremath{\mathrm{C/O}}\xspace}
\newcommand{\mic}{\ensuremath{\upmu\mathrm{m}}\xspace}
\newcommand{\as}{\hbox{$^{\prime\prime}$}\xspace}
\newcommand{\vsini}{\hbox{$v \sin i$}\xspace}
\newcommand{\loD}{\hbox{$\lambda/D$}\xspace}

\newcommand{\kms}{\ensuremath{\mathrm{km}\,\mathrm{s}^{-1}}\xspace}
\newcommand{\crires}{CRIRES\xspace}
\newcommand{\hirise}{HiRISE\xspace}

\newcommand{\formosa}{\texttt{ForMoSA}\xspace}

\usepackage[normalem]{ulem}

\begin{document}

\title{Eccentric and cool: A high-spectral-resolution view of\\51\,Eri\,b with VLT/HiRISE\footnote{Based on observations made with ESO Telescopes at the La Silla Paranal Observatory under programmes ID 112.25FU, 114.2712, and 115.284P}}
\titlerunning{51\,Eri\,b with VLT/\hirise}

\author{
    A.~Denis\inst{\ref{lam}}
    \and
    A.~Vigan\inst{\ref{lam}}
    \and
    G.~Chauvin\inst{\ref{mpia}}
    \and 
    A.~Lacquement\inst{\ref{ipag}}
    \and
    H.~Beust\inst{\ref{ipag}}
    \and 
    M.~Ravet\inst{\ref{lagrange}, \ref{ipag}}
    \and
    J.~Costes\inst{\ref{lam}}
    \and
    A.~Radcliffe\inst{\ref{lira}}
    \and
    S.~Martos\inst{\ref{ipag}}
    \and
    W.~Balmer\inst{\ref{johnhopkins}, \ref{stsci}}
    \and 
    T.~Stolker\inst{\ref{leiden}}
    \and
    P.~Palma-Bifani\inst{\ref{lira}}
    \and 
    B.~Rajpoot\inst{\ref{mpia}, \ref{lagrange}, \ref{lam}}
    \and
    A.~Simonnin\inst{\ref{lagrange}}
    \and
    S.~Petrus\inst{\ref{godar}}
    \and 
    T.~Forveille\inst{\ref{ipag}}
    \and 
    M.~Janson\inst{\ref{albanova}}
    \and
    A.~Reiners\inst{\ref{gott}}
    \and
    N.~Godoy\inst{\ref{lam}}
    \and
    D.~Cont\inst{\ref{munchen}, \ref{garching}}
    \and 
    L.~Nortmann\inst{\ref{gott}}
    \and 
    K.~Hoy\inst{\ref{portales}, \ref{esoc}, \ref{YEMS}}
    \and 
    A.~Zurlo\inst{\ref{portales}, \ref{YEMS}}
    \and
    H.~Anwand-Heerwart\inst{\ref{gott}}
    \and
    Y.~Charles\inst{\ref{lam}}
    \and
    A.~Costille\inst{\ref{lam}}
    \and
    M.~El Morsy\inst{\ref{sanantonio}}
    \and
    J.~Garcia\inst{\ref{lam}}
    \and
    M.~Houllé\inst{\ref{lam}}
    \and
    M.~Lopez\inst{\ref{lam}}
    \and
    G.~Murray\inst{\ref{durham}}
    \and
    E.~Muslimov\inst{\ref{lam},\ref{oxf}}
    \and
    G.~P.~P.~L.~Otten\inst{\ref{lam},\ref{taiwan}}
    \and
    J.~Paufique\inst{\ref{esog}}
    \and
    M.~Phillips\inst{\ref{exeter},\ref{ifa}}
    \and
    U.~Seemann\inst{\ref{esog}}
    \and
    A.~Viret\inst{\ref{lam}}
    \and
    G.~Zins\inst{\ref{esog}}
    }

\institute{
    Aix Marseille Univ, CNRS, CNES, LAM, Marseille, France \label{lam}
    \\ \email{\href{mailto:allan.denis@lam.fr}{allan.denis@lam.fr}}
    \and
    Max-Planck-Institut f¨ur Astronomie, K¨onigstuhl 17, 69117 Heidelberg, Germany \label{mpia}
    \and
    Univ. Grenoble Alpes, CNRS, IPAG, F-38000 Grenoble, France \label{ipag} 
    \and 
    Laboratoire J.L. Lagrange, Université Côte d'Azur, Observatoire de la côte d'Azur, CNRS, 06304 Nice, France \label{lagrange}
    \and 
    LIRA, Observatoire de Paris, Université PSL, Sorbonne Université, Université de Paris, 5 place Jules Janssen, 92195 Meudon, France \label{lira}
    \and
    Department of Physics \& Astronomy, John Hopkins University, 3400 N. Charles Street, Baltimore, MD 21218, USA \label{johnhopkins}
    \and
    Space Telescope Science Institute, 3700 San Martin Drive, Baltimore, MD 21218, USA \label{stsci}
    \and
    Leiden Observatory, Leiden University, Einsteinweg 55, 2333 CC Leiden, The Netherlands \label{leiden}
    \and 
    NASA-Goddard Space Flight Center, Greenbelt, MD 20771, USA \label{godar}
    \and 
    Department of Astronomy, Stockholm University, AlbaNova University Center, 10691 Stockholm, Sweden \label{albanova}
    \and
    Institute for Astrophysics und Geophysik, Georg-August University, Friedrich-Hund-Platz 1, 37077 Göttingen, Germany \label{gott}
    \and
    Universitäts-Sternwarte, Ludwig-Maximilians-Universität München, Scheinerstraße 1, 81679 München, Germany \label{munchen}
    \and
    Exzellenzcluster Origins, Boltzmannstraße 2, 85748 Garching, Germany \label{garching}
    \and 
    Instituto de Estudios Astrofísicos, Facultad de Ingeniería y Ciencias, Universidad Diego Portales, Av. Ejército 441, Santiago, Chile \label{portales}
    \and
    European Southern Observatory, Alonso de Cordova 3107, Vitacura, Santiago, Chile \label{esoc}
    \and
    Millennium Nucleus on Young Exoplanets and their Moons \label{YEMS}
    \and
    Department of Physics and Astronomy, University of Texas-San Antonio, San Antonio, TX, USA \label{sanantonio}
    \and
    Center for Advanced Instrumentation, Durham University, Durham, DH1 3LE, United Kindgom \label{durham}
    \and
    Dept. of Astrophysics, University of Oxford, Keble Road, Oxford, OX1 3RH, UK \label{oxf}
    \and
    Academia Sinica, Institute of Astronomy and Astrophysics, 11F Astronomy-Mathematics Building, NTU/AS campus, No. 1, Section 4, Roosevelt Rd., Taipei 10617, Taiwan \label{taiwan}
    \and
    European Southern Observatory (ESO), Karl-Schwarzschild-Str. 2, 85748 Garching, Germany \label{esog}
    \and
    Physics \& Astronomy Dpt, University of Exeter, Exeter, EX4 4QL, UK \label{exeter}
    \and
    Institute for Astronomy, University of Hawaii at Manoa, Honolulu, HI 96822, USA \label{ifa}
}

\date{Received 05 January 2025; accepted 31 January 2026}

\abstract 
{Discovered almost ten years ago, the giant planet 51 Eridani b is one of the least separated ($\approx$0.2\as) and faintest (J $\approx$ 19.74 mag) directly imaged exoplanets known to date.  Its atmospheric properties have been thoroughly investigated through low- and medium-resolution spectroscopic observations, enabling the robust characterization of the planet’s bulk parameters. However, the planet’s intrinsically high contrast renders high-resolution spectroscopic observations difficult, despite their potential to yield key measurements essential for a more comprehensive characterization. This study sought to constrain the planet’s radial velocity, enabling a full 3D orbital solution when integrated with previous measurements. We obtained four high-contrast, high-resolution (R $\approx$ 140\,000) spectroscopic datasets of the planet, collected over a two-year interval with the HiRISE visitor instrument at the VLT to derive the planet's radial velocity. Using self-consistent models of atmosphere, we were able to derive the radial velocity of the planet at each of the four epochs. These radial-velocity measurements were then used in combination with all existing relative astrometry in order to constrain the orbit of the planet.Our radial velocity measurements allowed us to break the degeneracy along the line of sight, making the unambiguous interpretation of the phase curve of the companion possible. We further constrained the orbital parameters, and particularly the eccentricity, for which we derive e = $0.55_{-0.07}^{+0.03}$. The relatively high eccentricity indicates that the system has experienced dynamical interactions induced by an external perturber. We place constraints on the mass and semimajor axis of a hypothetical, unseen outer planet capable of producing the observed high eccentricities.}

\keywords{
  Instrumentation: high angular resolution --
  Instrumentation: adaptive optics --
  Instrumentation: spectrographs --
  Techniques: high angular resolution --
  Techniques: spectroscopy --
  Infrared: planetary systems
}

\maketitle

\section{Introduction}
\label{sec:introduction}
\vspace{-0.2cm}
Since the first spatially resolved exoplanet detection \citep{chauvin_giant_2004}, direct imaging has enabled the discovery and characterization of companions at wide separations, probing the outer regions of exoplanetary systems \citep{Bowler2016}. Population-level studies have provided initial constraints on their demographics and formation pathways \citep{Nielsen2019, Vigan2021}, and, when combined with other detection techniques, direct imaging contributes to a global view of planetary system architectures across a broad range of masses and separations \citep{bonfils_harps_2013, batalha_exploring_2014, bowler_population-level_2020}.

Eccentricity, for instance, is an efficient indicator of planetary formation and dynamical evolution. While planets forming in protoplanetary disks are expected to form on circular orbits, they can develop nonzero eccentricities through dynamical processes such as planet–planet scattering \citep{ford_origins_2008}, Kozai–Lidov oscillations \citep{kozai_secular_1962, lidov_evolution_1962}, planet–disk interactions \citep{li_resonant_2023}, or stellar flybys \citep{pfalzner_trajectory_2024}. A population-level dichotomy exists between brown-dwarf and giant-planet companions, with brown dwarfs peaking at high eccentricities ($e \approx 0.6$–$0.9$) and giant planets typically exhibiting low eccentricities \citep{bowler_population-level_2020}. The giant planet 51\,Eridani\,b (hereafter 51\,Eri\,b) is an exception, as recent studies suggested a potentially high eccentricity \citep{rosa_updated_2020, dupuy_limits_2021, balmer_jwst-tst_2025}, which is indicative of past dynamical interactions.

Discovered in 2014 by the Gemini Planet Imager Exoplanet Survey \citep{macintosh_discovery_2015}, 51\,Eri\,b is orbiting an F0 star that is a member of the $\beta$~Pictoris moving group ($24 \pm 3$\,Myr; \citealt{Bell2015}). Updated stellar parameters yield $R_{\star} = 1.45 \pm 0.02,R_{\odot}$ and $M_{\star} = 1.550 \pm 0.006,M_{\odot}$, with an age of $23 \pm 1$\,Myr \citep{elliott_measuring_2024}. The system also hosts the wide, co-moving M-dwarf binary GJ\,3305\,AB at a projected separation of $\sim$2000 au \citep{feigelson_51_2006, kasper_novel_2007}.

The atmosphere of 51\,Eri\,b has been extensively characterized, revealing a cool ($\sim$650\,K), partially cloudy atmosphere with strong methane absorption features and disequilibrium chemistry \citep{samland_spectral_2017, brown-sevilla_revisiting_2023, balmer_jwst-tst_2025}. Using the Exo-REM forward model \citep{charnay_self-consistent_2018} and all available spectrophotometric observations from GPI, SPHERE, and JWST/NIRCam, \cite{balmer_jwst-tst_2025} derived $\Teff = 632 \pm 13$,K, a super-solar metallicity ($\met = 0.65 \pm 0.15$), $\co = 0.65 \pm 0.08$, $R_p = 1.3 \pm 0.03 \RJup$, and a low surface gravity ($\logg = 3.7 \pm 0.3$), confirming earlier indications of a metal-rich atmosphere \citep{samland_spectral_2017}. The orbit of 51\,Eri\,b has also been refined using astrometry and Gaia DR3 data, placing an upper limit on the planet mass (M$p < 9\,M{\mathrm{Jup}}$ at 3$\sigma$; \citealt{dupuy_limits_2021, balmer_jwst-tst_2025}) and revealing a highly eccentric orbit ($e = 0.57^{+0.03}_{-0.09}$; \citealt{balmer_jwst-tst_2025}).

In this paper, we present nearly two years of high-resolution ($R = 140,000$) direct spectroscopic observations of 51\,Eri\,b obtained with the \hirise instrument \citep{Vigan2024}, which combines SPHERE \citep{Beuzit2019} and \crires \citep{Dorn2023}. We describe the observations and data reduction in Sect.~\ref{sec:observations} and present the radial-velocity monitoring in Sect.~\ref{sec:RV}. In Sect.~\ref{sec:orbit}, we provide an update of the orbital solution realized by combining RVs with literature astrometry. We summarize our conclusions in Sect.~\ref{sec:conclusion}.

\vspace{-0.3cm}
\section{Observations and data reduction}
\label{sec:observations}

We observed the 51~Eri system at four epochs with VLT/\hirise (see Table~\ref{tab:observations}) following the same observing strategy for AF\,Lep\,b described in \cite{denis_characterization_2025}. For the first three epochs, sky backgrounds were acquired after the science exposures to subtract the leakage associated with the \hirise MACAO guide fiber \citep{Vigan2024}. After a technical intervention in January 2025, the leakage issue was fully resolved, and subsequent observations relied on internal daytime backgrounds.

All observations were obtained at low air masses ($<1.2$) under stable conditions, with a typical DIMM seeing between 0.5\as and 1.0\as. The end-to-end system transmission was measured on the star, with 95$^{\mathrm{th}}$ percentile values of $\sim$4\% for all epochs, which is consistent with instrumental expectations for bright targets \citep{Vigan2024}. This confirms accurate centering of the stellar PSF on the science fiber within the 0.2,\loD requirement \citep{ElMorsy2022}.

Standard \crires calibrations were acquired automatically the next morning based on the science observations of each night. These include dark, flat fields and wavelength-calibration files, as detailed in the \crires calibration plan. \hirise observations do not require any specific internal calibrations in addition to standard daily calibrations of the spectrograph.

Data reduction was performed using the public HiRISE pipeline \citep{hipipe2024}, which applies background and flat-field corrections, combines exposures, extracts the spectra, recalibrates the wavelength, and corrects for barycentric motion, while providing calibrated stellar and companion spectra,\footnotemark sky and instrumental response, and associated uncertainties. The pipeline is described in detail in \cite{denis_characterization_2025} and now includes an optimal extraction scheme \citep{horne_optimal_1986}, representing a significant improvement over earlier reductions.

\vspace{-0.3cm}
\section{Determination of the radial velocity of 51\,Eri\,b}
\label{sec:RV}

\subsection{Forward modeling analysis}
\label{subsec:forward_modeling}

To measure the radial velocity (RV) of the companion, we compared the data to self-consistent atmospheric models using the forward-modeling code \formosa \citep{petrus_medium-resolution_2021}. \formosa operates within a Bayesian framework and allows the joint analysis of spectroscopic and photometric data over a wide range of wavelengths and resolutions. Atmospheric model grids spanning effective temperature, gravity, metallicity, and related parameters are interpolated at each iteration, with optional modifications such as Doppler shifts and rotational broadening to be applied before computing the likelihood. The output consists of the best-fitting atmospheric parameters and their uncertainties.

We adopted the radiative–convective equilibrium model Exo-REM \citep{charnay_self-consistent_2018} recomputed at a spectral resolution of 200\,000, using low-resolution Exo-REM volume mixing ratio profiles in tandem with Exo\_k. This model is referred to as Exo-REM k26 (Radcliffe et al., submitted). A cloud-free model was used, along with the priors of \cite{balmer_jwst-tst_2025}. Given the high spectral resolution and narrow wavelength coverage, the data are expected to probe the upper atmospheric layers above the cloud base, as discussed in \citet{xuan_clear_2022} and \citet{denis_characterization_2025}. To compare the model to the data, we followed the framework described in \cite{denis_characterization_2025}.

The results are summarized in Table~\ref{tab:results}, which reports the planetary RV corrected for barycentric motion, the stellar RV \citep[$12.6 \pm 0.3$,\kms;][]{gontcharov_pulkovo_2006}, and the planetary \vsini. The \vsini measured at the fourth epoch is inconsistent with the other epochs, likely due to instrumental fringing (see Appendix~\ref{appendix:fringes}). We therefore relied on the first three epochs to estimate the rotational broadening and derive $\vsini = 9.9 \pm 2.3$\,\kms, keeping in mind that the fringing effect still affects these data. Assuming $\sin i = 1$ and adopting the planetary radius $R_p = 1.3 \pm 0.03,\RJup$ \citep{balmer_jwst-tst_2025}, we can derive an upper limit on the rotation period of the planet ($\sim 16.4 \pm 3.8$ h). We thus estimated $P_{\mathrm{rot}} \lesssim 24$ h at 2 $\sigma$. 

\begin{table}[!ht]
    \centering
    \caption{Results on 51\,Eri\,b for the our nights of observations.}
    \renewcommand{\arraystretch}{1.5}
    \begin{tabular}{lccc} \hline \hline
    Night & $\Delta$RV\tablefoottext{a} & \vsini & $R_{\mathrm{eff}}$ \tablefoottext{b} \\ 
          & [\kms] & [\kms] & \\
    \hline 
    2023-11-21 & 1.72$_{-1.99}^{+1.94}$ & 10.21$_{-3.21}^{+3.45}$ & 29\,363$_{-7\,416}^{+13\,625}$ \\
    2024-12-01 & 4.24$_{-2.01}^{+2.00}$ &  6.96$_{-3.00}^{+4.51}$ & 43\,074$_{-16\,987}^{+32\,631}$ \\
    2025-02-03 & 2.75$_{-2.90}^{+2.44}$ & 12.32$_{-4.19}^{+4.43}$ & 24\,334$_{-12\,541}^{+6\,436}$ \\ 
    2025-09-11 & 4.12$_{-0.90}^{+0.86}$ &  2.65$_{-1.04}^{+1.33}$ & 113\,129$_{-37\,804}^{+73\,077}$ \\
    \hline 
    \end{tabular}
    \label{tab:results}
    \tablefoot{
        \tablefoottext{a}{$\Delta$RV represents the difference between the RV of the companion and the RV of the star. 
        \tablefoottext{b}{The effective resolution, $R_{\mathrm{eff}}$, is computed as $c/\vsini$}.}
    }
\end{table}
\vspace{-0.2cm}

\subsection{Cross-correlation function}
\label{subsec:cross-correlation}

To assess the presence of individual molecules in the atmosphere, it is convenient to use a cross-correlation-function (CCF) analysis, in which we correlate molecular templates with the data. In the H band, at the temperature of 51\,Eri\,b, the dominant species are H$_2$O, CH$_4$, and CO. To generate the templates for these molecules, we used the pressure-temperature profiles and the volume-mixing ratio profiles of Exo-REM in tandem with Exo\_K. For the derivation of the CCF, we refer the reader to Appendix~\ref{appendix:ccf}.

There appears to be no correlation between the CCF and the derived RV uncertainty. For the first and third epochs, where detection confidence is highest, the derived RV uncertainties are more than twice those of the final epoch. Although it is natural to expect RV uncertainty to correlate with the CCF Signal-to-Noise (S/N), we are in a regime where resolution is also an important parameter. The lower vsini of the last epoch compared to the other epochs keeps the effective resolution of the template higher (see Table \ref{tab:results}). This results in sharper variations of the log likelihood near the RV of the planet and therefore less uncertainty in its determination.

\begin{figure}[!ht]
    \centering
    \includegraphics[width=0.9\linewidth]{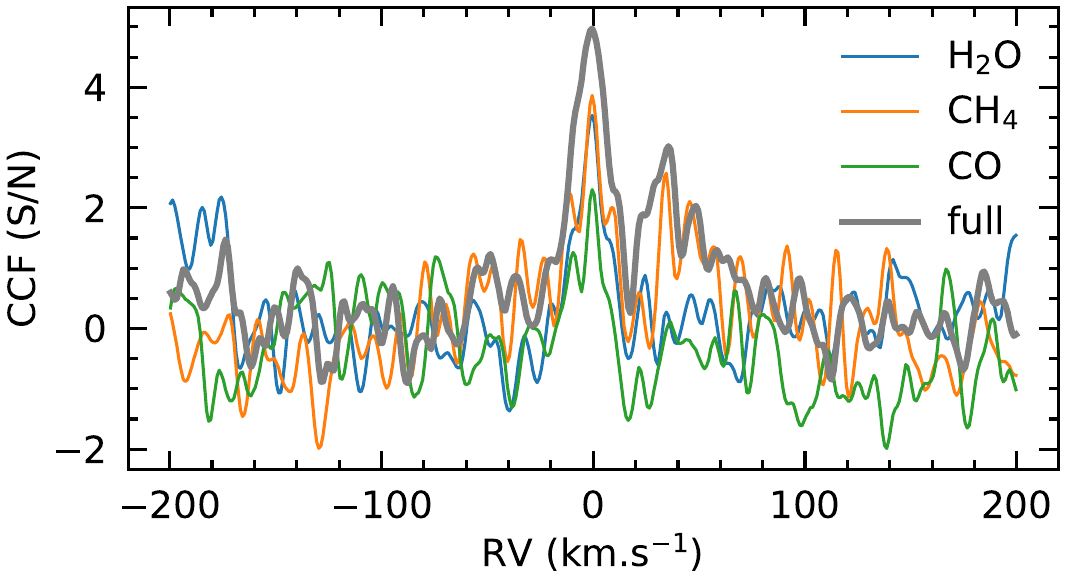}
    \caption{Combined CCF from the four nights. The combined CCF was computed by shifting the each individual CCF to an RV of 0 $\kms$ in order to account for the variations of the RV of the companion across the different epochs.}
    \label{fig:combined_ccf}
\end{figure}

Figure~\ref{fig:combined_ccf} shows the combined CCF from the four epochs. Before combining the four CCFs, we shifted each individual CCF to the rest frame of the planet to account for the variations in RV across the different epochs. In addition to the individual molecular templates, we also considered a full template including H$_2$O\,+\,CH$_4$\,+\,CO. In general, the molecular detections at individual epochs are marginal, but the combined CCF exhibits a reliable detection of H$_2$O and CH$_4$ and a tentative detection of CO, which confirms the recent findings of \cite{madurowicz_direct_2025}. 

\vspace{-0.5cm}
\section{Updated orbital solution}
\label{sec:orbit}

\begin{figure*}
    \centering
    \includegraphics[width=1\textwidth]{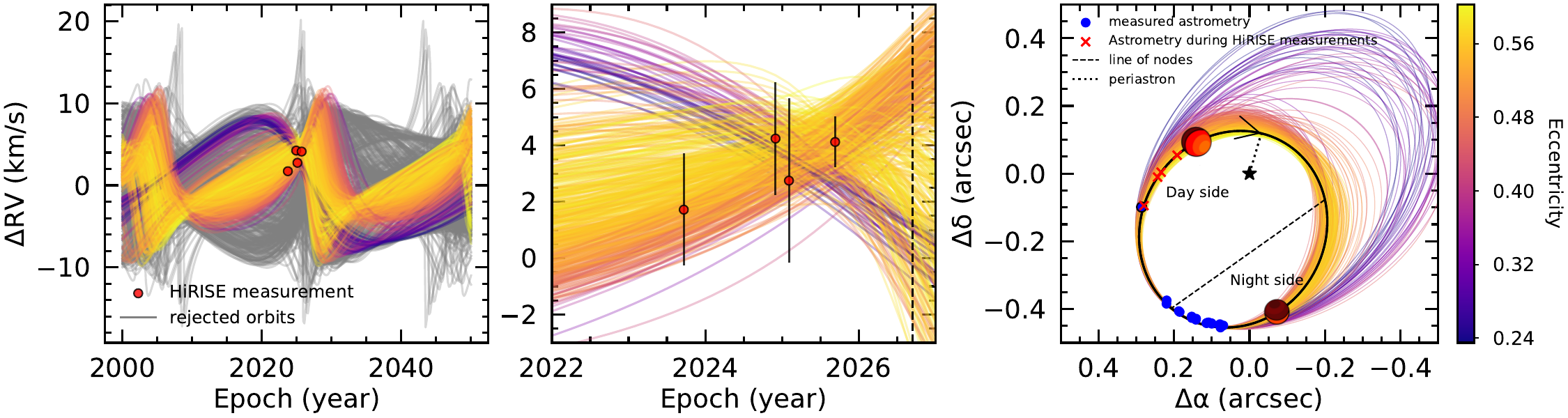}
    \caption{Orbit of 51\,Eri\,b. The left panel shows the predicted RV of the planet as a function of time. The HiRISE RV measurements are overplotted as red dots, along with the rejected orbits in grey from the addition of the RV measurements. The middle panel is a zoomed-in version of the left panel around our RV measurements. In particular, we see that a future RV measurement at the end of 2026 should enable us to reject the last low-eccentric orbital populations. This is illustrated by the dashed vertical line overplotted at the end of 2026. The right panel displays the astrometric orbit of the planet inferred, including the line of nodes and the phase of the planet. The red crosses represent the estimate of the position of the planet during the \hirise observations.}
    \label{fig:astrometry}
\end{figure*}

\begin{table}
    \centering
    \caption{Orbital parameters inferred in this work.}
    \begin{tabular}{lll}
        \hline
        \hline
        Parameter & Posterior & $\chi^2_{\mathrm{min}}$ \\
        \hline
        Semi-major axis $a$ (au) & $9.08_{-0.19}^{+0.57}$ & 8.91 \\
        Eccentricity $e$ & $0.55_{-0.07}^{+0.03}$ & 0.58 \\
        Inclination $i$ (\degree) & $159.24_{-11.48}^{+5.74}$ & 165 \\
        Argument of periastron $\omega$ (\degree) & $55.41_{-30.75}^{+51.00}$ & 84 \\
        Ascending node $\Omega$ (\degree) & $58.25_{-47.60}^{+80.80}$ & 69 \\
        period P (years) & $23.49_{-0.86}^{+1.74}$ & 22.76 \\
        Periastron passage $\tau$\tablefootmark{a} & $0.33_{-0.03}^{+0.02}$ & 0.33 \\
        \hline
    \end{tabular}
    \label{tab:obital_results}
    \tablefoot{
        \tablefoottext{a}{The periastron passage is computed as $\tau =\frac{t_{\mathrm{p}} - t_{\mathrm{ref}}}{P}$, where t$_{\mathrm{ref}}$ is the reference time at epoch J2020.0 ($t_{\mathrm{ref}}$ = 58849 MJD) and t$_{\mathrm{p}}$ is the time of periastron.}
        }
\end{table}

To update the orbit of 51 Eri b, we compiled relative astrometric data from Gemini/GPI and Keck/NIRC2 \citep{macintosh_discovery_2015, rosa_updated_2020}, VLT/SPHERE \citep{maire_hint_2019}, and JWST/NIRCam \citep{balmer_jwst-tst_2025}. We used the Markov chain Monte Carlo (MCMC) framework designed by \cite{beust_orbital_2016}, which makes use of universal Keplerian variables with Stumpff functions \citep{danby_solution_1987}. This formulation achieves more efficient convergence in the case of astrometric orbits sampled over a small orbital arc, as in the present case. The MCMC was run using eight walkers. The burn-in and convergence was assessed using Gelman-Rubin statistics \citep{gelman_bayesian_2003, ford_improving_2006}. Before beginning the inference of the chains, we required that $\hat{R}(z)<1.1$ and $\hat{T}(z)>100$ (see Sects. 3.3 and 3.4 of \citealt{ford_improving_2006}). Then, we let the chains be filled until $\hat{R}(z)<1.01$ and $\hat{T}(z)>1000$. The filled chains constitute the output of the orbital fit.

We did not directly fit for classical orbital elements: the semimajor axis, $a$; the eccentricity, $e$; the inclination, $i$; the argument of periastron, $\omega$; the longitude of the ascending node, $\Omega$; and the time of periastron passage, $t_p$. \cite{ford_improving_2006} suggested that working with combinations of these elements allows an optimization of the Markov chains.  We specifically fit for $\ln q$, $e\cos(\Omega+\omega)$, $e\sin(\Omega+\omega)$, $sin(i/2)\cos\Omega$, $\sin(i/2)\sin\Omega$, $\ln P$, and $s$. Here, $q$ is the periastron, $P_q$ is the orbital period of a circular orbit with radius $q$, and $s$ is the universal variable (a reformulation of the eccentric anomaly; \citealt{danby_solution_1987}) at a reference epoch. Note that $P$ can also be defined as $P_q=2\pi qv_q$, where $v_q$ is the velocity at periastron. From any solution with those fit parameters, the classical elements can be derived unambiguously. 

The updated orbit and associated parameters are presented in Fig.~\ref{fig:astrometry} and Table~\ref{tab:obital_results}. The full posterior distributions of orbital parameters are presented in Fig.~\ref{fig:corner}. The left panel of Fig.~\ref{fig:astrometry} shows the predicted RV differences between the planet and the star as a function of time. The middle panel shows a zoomed-in view centered on the HiRISE measurements, and the right panel displays the astrometric orbit of the companion. These plots were generated using 500 orbits randomly drawn from the posterior distributions.

The inclusion of our RV measurements rules out a significant number of orbital solutions, as can be seen in the left panel of Fig.~\ref{fig:astrometry}. The middle panel shows that a future RV measurement of the planet at the end of 2026 will potentially completely reject the last population of low-eccentricity orbits and place even tighter constraints on the eccentricity. As illustrated in Fig.~\ref{fig:corner}, adding the HiRISE RV measurements significantly improves the orbit. We now estimate the eccentricity to lie within the 0.2--0.7 range at $5\sigma$, with a median at $e = 0.55$, completely ruling out very eccentric orbits, above 0.7.

The HiRISE measurements also allowed us to break the typical degeneracy arising with the argument of periastron, $\omega$, and the longitude of ascending node, $\Omega$, when using only astrometric data to fit orbits of exoplanets. This is because the computation of the astrometry of the planet from the orbital parameters is invariant to the transformation ($\omega'=\omega+\pi$, $\Omega'=\Omega+\pi$), typically producing double-peaked structures in the posterior distributions of these parameters. As a consequence, including our RV measurements enabled the determination of the direction of the motion of the companion along the line of sight. This allowed us to compute a phase curve of the companion (see the right panel of Fig.~\ref{fig:contrast_curve}), which is essential for planning follow-up observations using reflected-light imaging or spectroscopy with future instrumentation. This is particularly pertinent in this case, since the reflected-light contrast is predicted to be higher than the thermal-light contrast for this planet, mostly under favorable phase functions (see Appendix.~\ref{appendix:contrast}).

\vspace{-0.3cm}
\section{Discussion and conclusions}
\label{sec:conclusion}

Our very-high-resolution H-band spectroscopic observations ($\lambda = 1.4$–$1.8$,\mic, $R \approx 140,000$) of 51\,Eri\,b led to detections at four epochs over roughly two years, providing new orbital constraints. In particular, we broke the degeneracy in the direction of motion along the line of sight, enabling the determination of the planet’s phase curve, which is crucial for future reflected-light observations. Most low-eccentricity orbits were ruled out, further confirming the eccentric nature of the planet. The combined CCF shows marginal detections of H$_2$O, CH$_4$, and CO, consistent with the faintness of the planet, with the weak CO signal likely due to its minor contribution in the H band at this temperature. Finally, from the projected rotational-velocity estimate of the planet, we derived an upper limit for its rotation period (P $\lesssim$ 24 hours at 2 $\sigma$). Future complementary observations will be required to strengthen molecule detections and give a more robust upper-limit estimate of the rotation period of the planet.

The implications of the planet’s high eccentricity have been extensively discussed in previous studies \citep{maire_hint_2019, dupuy_limits_2021, balmer_jwst-tst_2025}. Dynamical interactions with a massive inner companion could in principle reproduce the observed eccentricity, but the required mass would exceed SPHERE detection limits \citep{maire_hint_2019}. An alternative scenario involves secular resonances with the massive circumstellar disk, which can excite planetary eccentricities up to $\sim$0.6, even for modest disk eccentricities \citep{li_resonant_2023}.
Kozai–Lidov oscillations induced by a distant third body, such as the stellar companion GJ\,3305\,AB, can also generate large eccentricities \citep{kozai_secular_1962, lidov_evolution_1962}. However, the wide separation of GJ\,3305\,AB implies a Kozai–Lidov timescale longer than the age of the system, making this scenario unlikely \citep{montet_dynamical_2015}, except in the case of a very high stellar companion eccentricity ($\gtrsim 0.9$; see Appendix~\ref{appendix:Kozai-Lidov}). Kozai–Lidov oscillations with an undetected low-mass outer companion remain possible; given SPHERE/IRDIS detection limits \citep{chomez_sphere_2025}, such a companion would likely be Jovian or sub-Jovian ($\lesssim 1$–$2,\MJup$). Continued astrometric and RV monitoring will further constrain this scenario \citep{lacour_mass_2021}.

If Kozai–Lidov oscillations can be ruled out, past dynamical interactions, such as a stellar flyby \citep{pfalzner_trajectory_2024}, offer another possible explanation. Further N-body simulations and searches for potential stellar perturbers in Gaia data, as performed for the HD~106906 system \citep{de_rosa_near-coplanar_2019}, will provide additional insight into the formation and dynamical evolution of the 51~Eri system.

\begin{acknowledgements}
    This project has received funding from \emph{Agence Nationale de la Recherche} (ANR) under grant ANR-23-CE31-0006-01 (MIRAGES). The \hirise instrument has been developed with funding from the European Research Council (ERC) under the European Union's Horizon 2020 research and innovation programme, grant agreements No. 757561 (\hirise) and 678777 (ICARUS), from the \emph{Commission Spécialisée Astronomie-Astrophysique} (CSAA) of CNRS/INSU, and from the \emph{Action Spécifique Haute Résolution Angulaire} (ASHRA) of CNRS/INSU co-funded by CNES, and from Région Provence-Alpes-Côte d'Azur under grant agreement 2014-0276 (ASOREX).
    This research has made use of computing facilities operated by CeSAM data center at LAM, Marseille, France.
    S. Petrus was supported by an appointment to the NASA Postdoctoral Program at the NASA-Goddard Space Flight Center, administered by Oak Ridge Associated Universities under contract with NASA. 
    N. Godoy acknowledges funding from the European Union (ERC, ESCAPE, project No 101044152).
\end{acknowledgements}

\bibliographystyle{aa}
\vspace{-0.5cm}
\bibliography{51Eri}

\appendix
\label{Appendix}

\section{Observations of 51\,Eri\,b}
\label{appendix:observations}

A summary of the HiRISE observations of 51\,Eri\,b is presented in Table \ref{tab:observations}.

\begin{table*}[!ht]
  \caption[]{51 Eri observations.}
  \renewcommand{\arraystretch}{1.2}
  \label{tab:observations}
  \centering
  \begin{tabular}{ccccccccc}
    \hline\hline
    UT date    & Setting  & Integration time & Integration time & Seeing        & Transmission\tablefoottext{a} & Angular separation\tablefoottext{b} \\
               &          & on star          & on companion \\
               &        & [min]                     & [min]                         & [\as]         & [\%]                          & [mas] \\
    \hline
    2023-11-21 & H1567    & 4                        & 100                           & 0.6           & 4.0                           & 295 \\
    2024-12-01 & H1567    & 4                        & 120                           & 0.6 - 0.7     & 3.9                           & 244 \\
    2025-02-03 & H1567    & 12                       & 120                           & 0.5 - 0.9     & 3.9                           & 234 \\ 
    2025-09-11 & H1567    & 4                        & 120                           & 0.5 - 0.8     & 3.7                           & 200 \\
    \hline
  \end{tabular}
  \tablefoot{
    \tablefoottext{a}{The transmission is measured on the star. The value provided corresponds to the 95$^{th}$-percentile of the transmission computed over the whole spectral range covered with HiRISE.} \tablefoottext{b}{The angular separation was computed from the orbital fitting (see Sect.~\ref{sec:orbit}).}
    }
\end{table*}

\section{Considering whether the results of the fourth epoch are biased}
\label{appendix:fringes}

\begin{figure}[!ht]
    \centering
    \includegraphics[width=\linewidth]{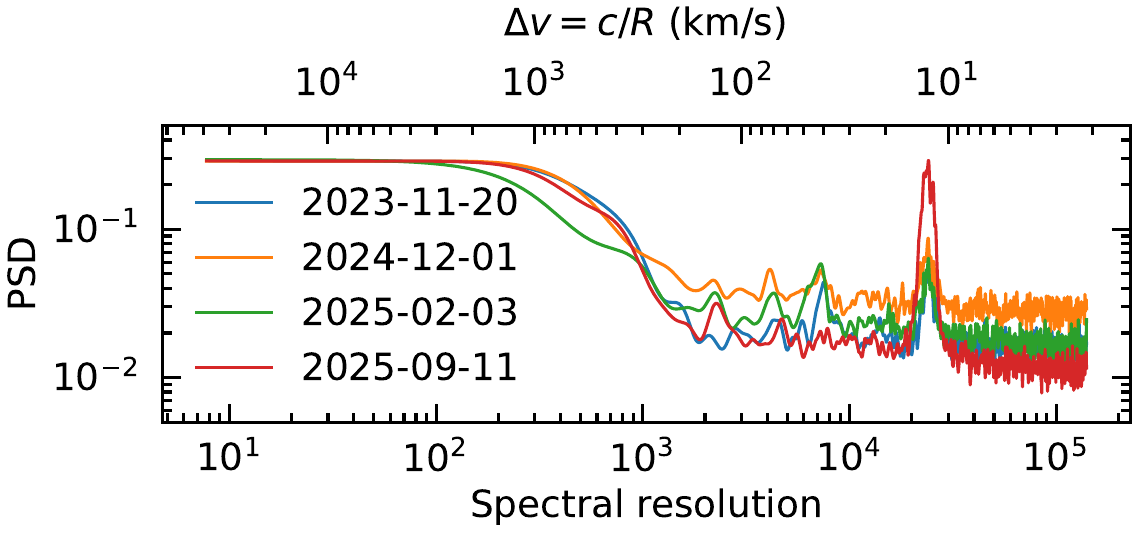}
    \caption{PSD of the data filtered from the stellar speckles estimate of our four epochs. The PSD was lowpass-filtered in order to focus on its low-frequency content. A secondary axis is displayed at the top, which corresponds to an equivalent velocity. A peak is visible around a resolution of 25\,000 for each epoch, which corresponds to $\sim$12\,\kms. This peak is particularly prominent in the data from the last epoch, when the planet was closer to the host star.}
    \label{fig:fringes}
\end{figure}

In Table~\ref{tab:results}, we see that the \vsini of the last epoch is particularly inconsistent with the other epochs. A possible explanation of this bias is the presence of spectral fringes in the data due to the dichroic filter present in \hirise \citep[see Fig.~1 of][]{ElMorsy2022}. This dichroic filter is a plane-parallel piece of glass with anti-reflection coatings, which has the unfortunate property of acting as a Fabry-Pérot cavity and causing oscillations as a function of wavelength in the stellar speckles signal. This is identified in the power-spectral density (PSD) of the data filtered from the stellar speckle estimates, which is plotted for our four epochs in Fig.~\ref{fig:fringes}. The peaks located at a spectral resolution of $\sim$25\,000 correspond to the fringes. The equivalent wavelength in the velocity space is $\approx$12\,\kms. It means that the fringes induce a periodic signal in the spectrum whose peaks are separated by $\sim$12\,\kms in the velocity dimension. This translates into correlated structures in the CCF at multiples of $\sim$12\,\kms.

\begin{figure}
    \centering
    \includegraphics[width=\linewidth]{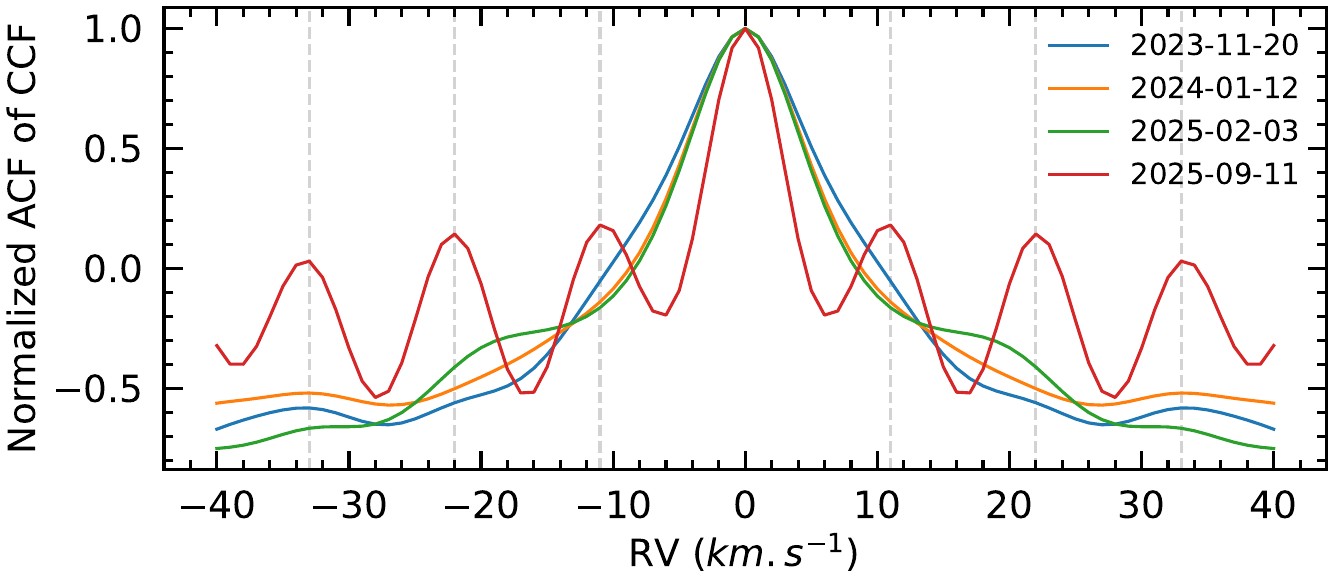}
    \caption{Normalized ACFs of the individual CCFs. The first three epochs show highly consistent ACFs, whereas the last epoch shows deeply correlated structures in the CCF at multiples of 11\,km/s. The vertical dashed lines are plotted at multiples of $\pm$ 11 \kms to highlight the correlated structures.}
    \label{fig:acf}
\end{figure}

To highlight these correlated structures, it is convenient to compute the autocorrelation function (ACF) of the CCF. Figure~\ref{fig:acf} presents the individual ACFs of the 4 CCFs. The first three epochs are very consistent. The last epoch, however, exhibits prominent correlated structures at velocities multiples of $\sim$ 11 \kms, which is consistent with the previous analysis. It also shows that the fringes cause a decrease in the full width at half-maximum (FWHM) of the ACF, compared to the other epochs. It means that the fringes bias the FWHM of the CCF to lower values. Since we expect the FWHM of the CCF and the \vsini to be correlated \citep[see][for example]{brady_measuring_2023}, we can conclude that the fringes likely bias the \vsini estimate of the last epoch towards lower values. However, they likely do not bias the RV as much. Provided that the template correlates better with the data than the fringes do, the CCF still provides a correct measurement of the RV of the planet.

This effect is actually quite common and has been identified in other contexts such as in the KPIC instrument \citep{horstman_fringing_2024} or in the JWST/MIRI medium-resolution spectroscopy mode \citep[][]{martos_combining_2025}. This systematic noise can become a significant limitation in the detection of the planet and its intensity is proportional to the intensity of the stellar light in the speckle field at the location of the companion. If we compute the angular separation of the companion at each of our observing epochs (see Table~\ref{tab:observations}), we see that the planet was at its closest to the host star during the last epoch. And indeed looking at the PSD in Fig.~\ref{fig:fringes}, this same epoch is also the one where the peak corresponding to the fringes is the strongest. 

A potential solution would to be include a model of the fringing signal, as in \cite{horstman_fringing_2024}. However, this is beyond the scope of the present paper and we consider the \vsini estimate of the fourth epoch to be unreliable.

\section{Method for computing the cross-correlation functions}
\label{appendix:ccf}

\begin{figure*}[!ht]
    \centering
    \includegraphics[width=\textwidth]{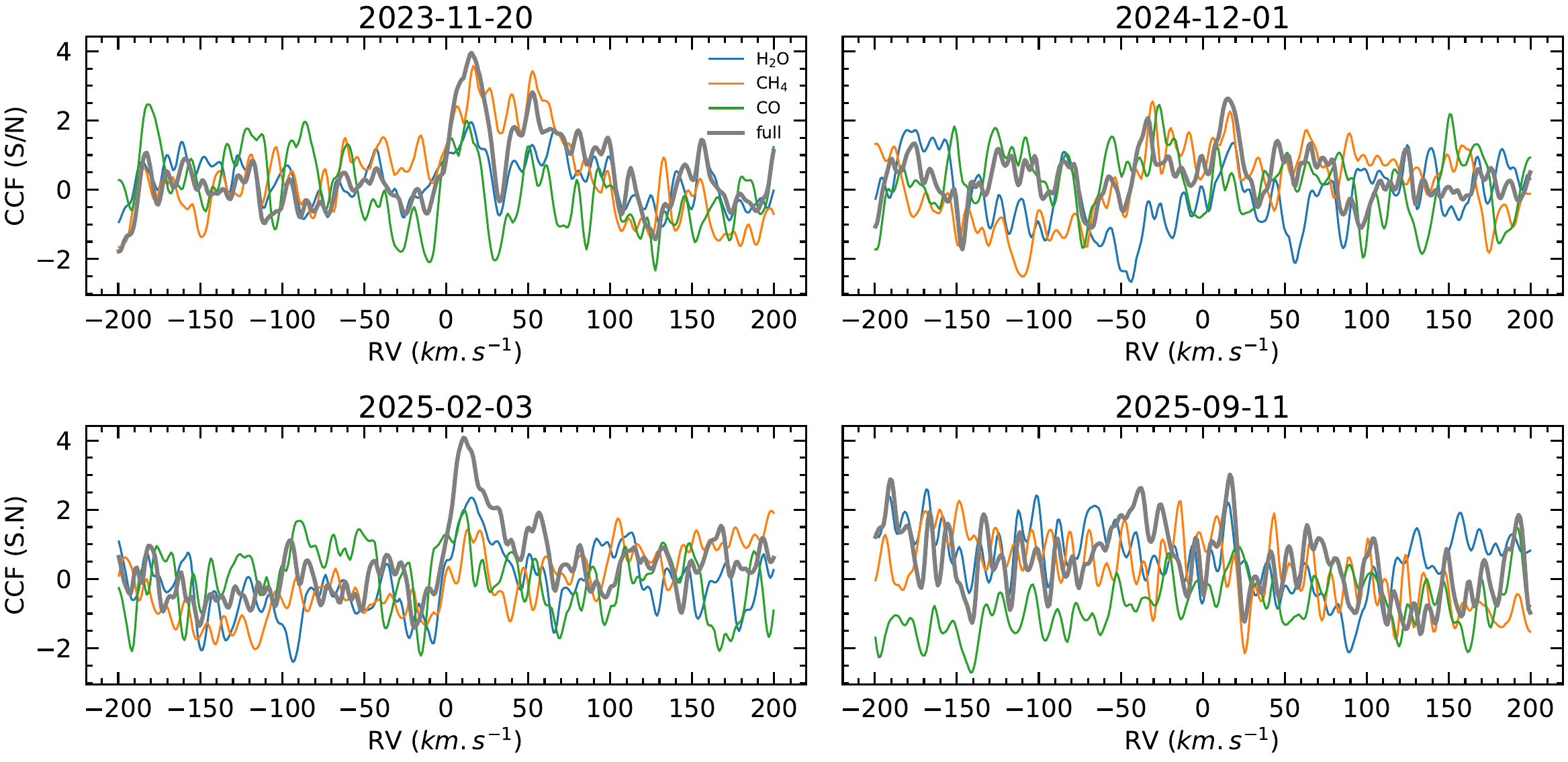}
    \caption{Individual CCFs for the four epochs.}
    \label{fig:ccf}
\end{figure*}

To compute the cross-correlation functions, we use the formalism defined in \cite{zucker_cross-correlation_2003}. The cross-correlation function is defined as 
\begin{equation}
    CCF(RV) = \frac{\sum_id_{HF}(i)t_{HF}^{RV}(i)}{\sqrt{\sum_id_{HF}(i)^2}\sqrt{\sum_it_{HF}^{RV}(i)^2}}.
\end{equation}
In this equation, d$_{HF}$ is the data filtered from the star speckles contamination and t$_{HF}^{RV}$ is the high-passed filtered template Doppler-shifted by a value of $RV\,\kms$. We refer to \citet{bidot_exoplanet_2024}, \citet{landman_beta_2024}, and \citet{denis_characterization_2025} for the definition of these terms. The signal-to-noise ratio (S/N) was derived computing the standard deviation in two windows 200 \kms away from the main peak. Figure~\ref{fig:ccf} shows the resulting CCF for the four epochs. The detections are weak, which is not surprising given the faintness of the planet in the H band.

\section{Orbit posterior distribution}
\label{appendix:posterior_distribution}

\begin{figure*}[!ht]
    \centering
    \includegraphics[width=0.8\textwidth]{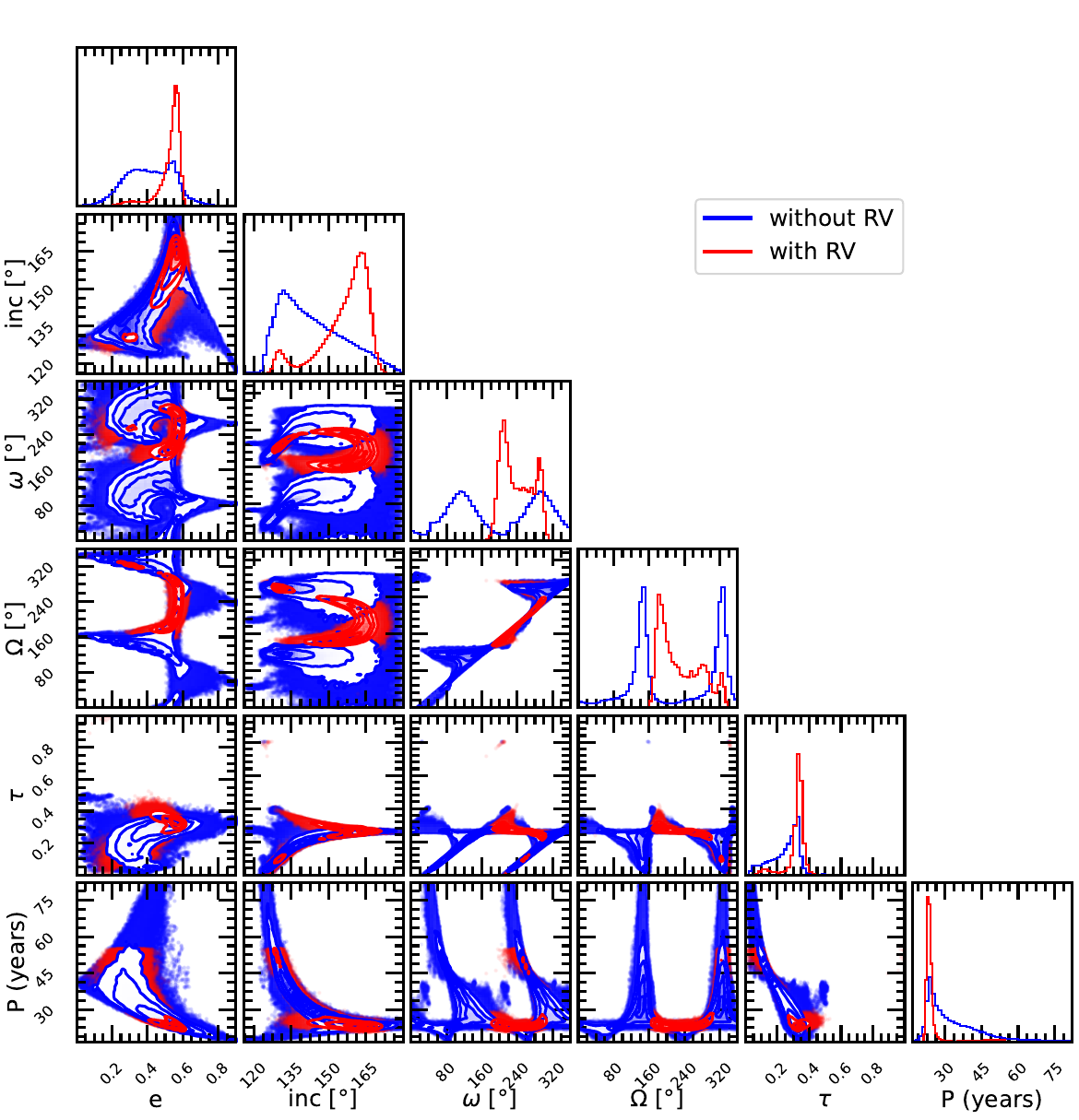}
    \caption{Posterior distribution of the inferred parameters of 51\,Eri\,b. The posterior distribution in red was obtained including all available relative astrometry measurements. The posterior distribution in blue includes the HiRISE RV measurements.}
    \label{fig:corner}
\end{figure*}

Figure~\ref{fig:corner} presents the posterior distribution of the orbital parameters of 51\,Eri\,b, without our RV measurements (in red) and with our RV measurements (in blue). The inclusion of the HiRISE RV measurements brings significant improvements to the orbital parameters. It also breaks the degeneracy between particular values of $\omega$ and $\Omega$ typically arising with orbital fittings using only astrometric data points.

\section{Contrast in reflected and emitted light}
\label{appendix:contrast}

In order to determine whether the planet could be a target of interest for reflected-light imaging and spectroscopy, it is crucial to establish whether the reflected-light contrast overpowers the thermal light contrast. We compared the predicted reflected-light contrast with the thermal light contrast within the visible spectrum. The reflected-light contrast was computed using the following formula:
\begin{equation}
    C = f(\phi) \, A \, \left(\frac{R}{a}\right)^2,
\end{equation}
with $R$ the radius of the planet, $a$ its semi-major axis, $A$ its albedo, and $f$ the phase function. We considered three different values for the phase function to show the impact of the phase of the planet. For $A$, we considered a value of 0.5, similar to that of Jupiter in visible \citep{karkoschka_methane_1998}. The values of $a$ and $R$ have been determined from the orbital and atmospheric fit, respectively. 

To compute the thermal light contrast in visible of the planet, we first compared the low-resolution spectro-photometric observations of 51\,Eri\,b with the self-consistent model Exo-REM in ForMoSA. We used the Y/J/H1-band band spectra from VLT/SPHERE \citep{brown-sevilla_revisiting_2023}, the Keck/GPI K1 and K1-bands spectroscopy and the Keck/NIRC2 L and M-bands photometry \citep{rajan_characterizing_2017} and the JWST/NIRCam F410M photometry \citep{balmer_jwst-tst_2025}. The results are consistent with the ones presented in \cite{balmer_jwst-tst_2025}. Then, because the Exo-REM model is not extended in the visible light range, we computed the flux in the visible light range using petitRADTRANS \citep{molliere_petitradtrans_2019} with the best Exo-REM PT profile and volume mixing ratio inferred from the forward modeling. 

\begin{figure*}[!ht]
    \centering
    \includegraphics[width=\textwidth]{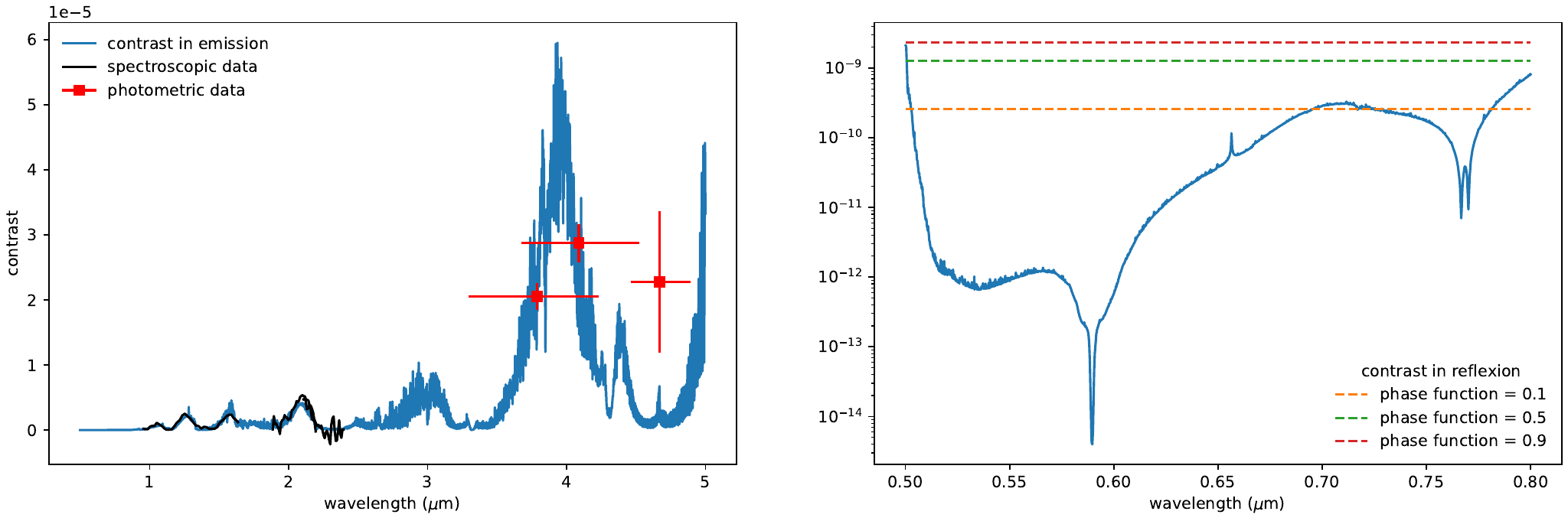}
    \caption{Contrast in thermal and reflected light. Left: Contrast curve in thermal light predicted for the companion between 0.5 and 5 $\mu$m. We overplotted the Y/J/H1-band SPHERE data from \cite{brown-sevilla_revisiting_2023}, the Keck/GPI K1- and K2-band spectroscopy and the Keck/NIRC2 L and M-bands photometry from \cite{rajan_characterizing_2017}, and the JWST/NIRCam F410M-band photometry from \cite{balmer_jwst-tst_2025}. Right: Zoomed-in version in the visible light range. We overplotted estimates of the reflected light contrast of the planet under different phase function scenarios, considering an albedo of 0.5.
    \label{fig:contrast_curve}}
\end{figure*}

The results are presented in Fig.~\ref{fig:contrast_curve} (right). For favourable phase angles of the planet, we see that the predicted reflected-light contrast is above the thermal light contrast. For very unfavourable phase angles of the planet, there is some regions where the thermal light is above the predicted reflected light, but between 0.5 and 0.7\,\mic, the reflected-light contrast is still predominant compared to the thermal light contrast. These conclusions do not dramatically change even when considering a very low albedo, such as 0.1.

\begin{figure*}[!ht]
    \centering
    \includegraphics[width=\textwidth]{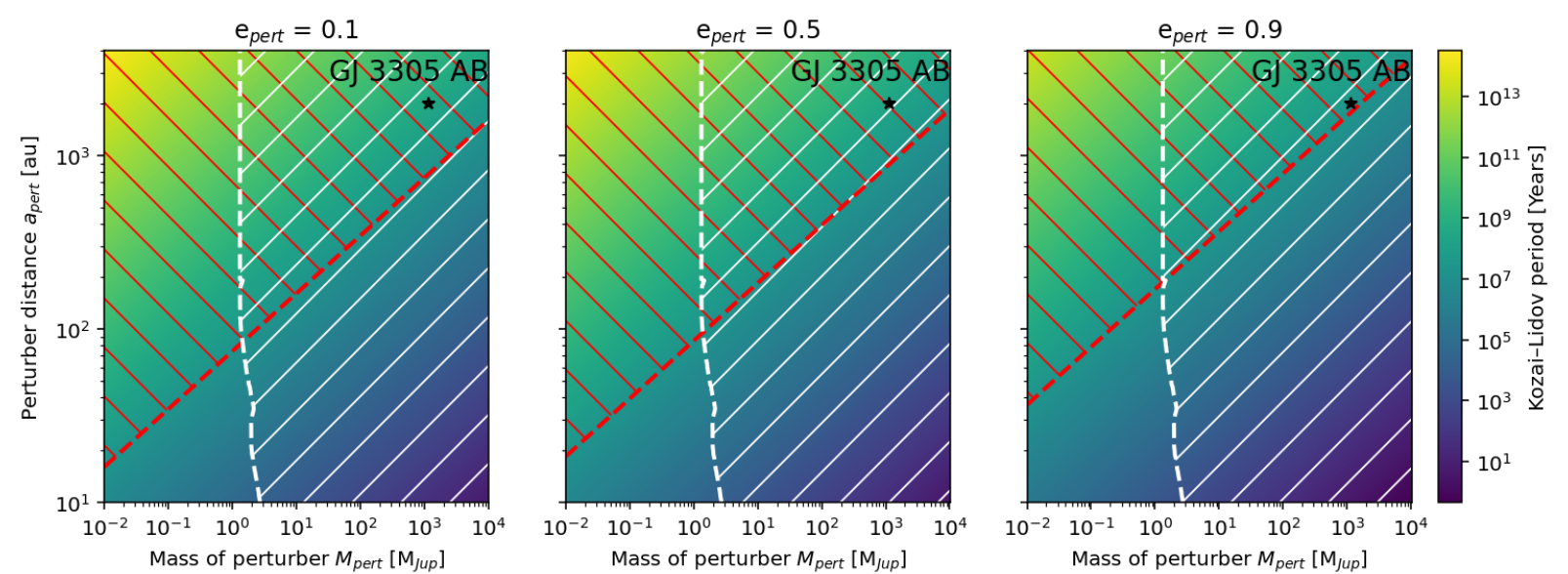}
    \caption{Kozai–Lidov oscillation timescale as a function of the perturber mass and semi-major axis for different eccentricities. The SPHERE/IRDIS detection limits are shown in white; perturbers located to the right of this curve (white hatched region) would be detectable by SPHERE. The red curve indicates the age of the system, with the red hatched region corresponding to Kozai–Lidov timescales longer than the system age. The position of the stellar binary companion GJ\,3305\,AB is also indicated.}
    \label{fig:Kozai-Lidov}
\end{figure*}

\section{Kozai-Lidov oscillations with 51\,Eri\,b}
\label{appendix:Kozai-Lidov}

In the Kozai-Lidov scenario, the orbit of a planet is perturbed by a distant third body, which can lead to oscillations in the inclination and eccentricity of the inner companion. The timescale of such oscillations is on the order of magnitude of \citep{ford_origins_2008}
\begin{equation}
    P_{\mathrm{Kozai}} \approx P_{\mathrm{planet}}\frac{M_{\star}}{M_{\mathrm{pert}}}\left(\frac{a_{\mathrm{pert}}}{a_{\mathrm{planet}}}\right)^3\left(1 - e_{\mathrm{pert}}^2\right)^{3/2}.
\end{equation}
In this equation, $P_{\mathrm{Kozai}}$ is the period of the Kozai-Lidov oscillations, $M_{\star}$, the mass of the host star, $M_{\mathrm{pert}}$, $a_{\mathrm{pert}}$ and $e_{\mathrm{pert}}$ the mass, semi-major axis and eccentricity of the distant perturber, and $P_{\mathrm{planet}}$ and $a_{\mathrm{planet}}$ the period and semi-major axis of the inner planet. We computed the period of the Kozai-Lidov oscillations for different values of $M_{\mathrm{pert}}$, $a_{\mathrm{pert}}$ and $e_{\mathrm{pert}}$ using the posterior distributions of $P_{\mathrm{planet}}$ and $a_{\mathrm{planet}}$ inferred in this study. Figure~\ref{fig:Kozai-Lidov} shows the results of this analysis. The red curve shows the age of the system. The parameter space for Kozai-Lidov oscillations on a timescale longer than the age of the system lies above this curve. Therefore, this parameter space is not applicable. The white curve represents the SPHERE detection limits from \cite{chomez_sphere_2025}. Hence, the range of possible solutions for the mass and semi-major axis of the unseen perturber lies between these 2 curves. The binary stellar companion GJ 3305 AB is overplotted. We considered the instantaneous sky-projected separation as the semi-major axis of the system and  ($\approx$ 2000 AU, \citealt{feigelson_51_2006}) and the dynamical mass estimate of the binary ($\approx$ 1.1 M$_{\odot}$, \citealt{montet_dynamical_2015}). If the binary stellar companion has a high eccentricity ($\gtrsim$ 0.9), it could potentially be responsible for the orbit of 51 Eri b, especially since, with such a high eccentricity, the semi-major axis could be significantly lower than the value of 2000 AU used in this analysis, lowering the Kozai-Lidov period. Indeed, the maximum possible separation of the stellar companion is $r_{apastron} = (1+e)*a$. This provides a lower limit for the semi-major axis of the stellar companion, i.e. $a > \frac{r_{apastron}}{1+e}$. For extreme values of the eccentricity, the minimum possible semi-major axis is $a_{min} = \frac{r_{apastron}}{2}$. This yields a range of possible values for the semi-major axis of the stellar companion, which can be as low as half of the value of the instantaneous sky-projected separation. This is a very unlikely scenario, however. Additional constrains on the possibility of GJ 3305 AB being in Kozai-Lidov resonance with 51\,Eri\,b could be brought by a direct measurement of the RV of the stellar companion, as it would help constrain both the inclination and the longitude of ascending node of the stellar companion. The Kozai-Lidov resonance requires the mutual inclination I between the 2 objects in resonance satisfies \citep{kozai_secular_1962, lidov_evolution_1962}\begin{equation}
    39\degree < I < 141\degree.
\end{equation}
From the RV measurement of the stellar companion, we can better constrain its inclination and longitude of ascending node. Then, the mutual inclination between 51\,Eri\,b and GJ 3305 AB is computed as \citep{murray_solar_2000}
\begin{equation}
\cos I = \cos i_1\cos i_2 + \sin i_1 \sin i_2 \cos(\Omega_1 - \Omega_2),
\end{equation}
with i$_1$, i$_2$, $\Omega_1$, $\Omega_2$ the inclinations and longitude of ascending node of the two companions.

\end{document}